\documentclass[prl,aps,twocolumn,nofootinbib,,superscriptaddress]{revtex4-1}
\usepackage{amsmath}
\usepackage{amsfonts}
\usepackage{amssymb}

\newcommand{\up}{\uparrow}
\newcommand{\down}{\downarrow}
\newcommand{\vect}[1]{\mathbf{#1}}
\usepackage{amsmath}
\usepackage{amsfonts}
\usepackage{graphicx}
\usepackage{verbatim} 
\newcommand{\Z}{\mathbb{Z}}

\begin{document}
\title{Interaction effects on 1D fermionic symmetry protected topological phases}

\author{Evelyn Tang}
\affiliation{Department of Physics, Massachusetts Institute of Technology, Cambridge, MA 02139}
\author{Xiao-Gang Wen}
\affiliation{Department of Physics, Massachusetts Institute of Technology, Cambridge, MA 02139}
\affiliation{Perimeter Institute for Theoretical Physics, Waterloo, Ontario, N2L 2Y5 Canada}

\date{Mar, 2012}

\begin{abstract}
In free fermion systems with given symmetry and dimension, the possible
topological phases are labeled by elements of only three types of Abelian
groups, $\Z_1$, $\Z_2$, or $\Z$.  For example non-interacting 1D fermionic
superconducting phases with $S_z$ spin rotation and time-reversal symmetries
are classified by $\Z$.  We show that with weak interactions, this
classification reduces to $\Z_4$.  Using group cohomology, one can additionally
show that there are only four distinct phases for such 1D superconductors even
with strong interactions. Comparing their projective representations, we find
all these four symmetry protected topological phases can be realized with free
fermions. Further, we show that 1D fermionic superconducting phases with $Z_n$
discrete $S_z$ spin rotation and time-reversal symmetries are classified by
$\Z_4$ when $n=$ even and $\Z_2$ when $n=$ odd; again, all these strongly
interacting topological phases can be realized by non-interacting fermions.
Our approach can be applied to systems with other symmetries to see
which 1D topological phases can be realized by free fermions.

\end{abstract}

\maketitle

Symmetry protected topological (SPT) phases\cite{GW0931,CGW1038} are short
range entangled states with symmetry protected gapless edge
excitations.\cite{PBT1039,CGW1107,SPC1032,CLW1141,CGL1172,GW1248} The Haldane phase
on a spin-1 chain\cite{H8364,AKL8877} and 2D/3D topological
insulators\cite{KM0501,BZ0602,KM0502,MB0706,FKM0703,QHZ0824} are examples of
SPT states.  Using K-theory or topological terms, all free-fermion SPT phases
can be classified\cite{K0886,SRF0825} for all 10 Altland-Zirnbauer symmetry
classes\cite{AZ9742} of \emph{single-body} Hamiltonians.  It turns out that
different free fermion SPT phases are described by only three types of Abelian
groups, $\Z_1$, $\Z_2$, or $\Z$. 

With interactions the classification is more varied, however we must first
describe the symmetry differently. Instead of specifying the symmetry of
single-body Hamiltonians, we treat the free fermion systems as many-body
systems and specify the \emph{many-body} symmetry of their \emph{many-body}
Hamiltonians.  Only in this case can we accurately add interaction terms to the
many-body Hamiltonian that preserve the many-body symmetry, and study their
effect on the SPT phases of free fermions. A classification of various
free-fermion gapped phases given their many-body symmetry can be found in Ref.
\onlinecite{W1103}.

Fidkowski and Kitaev (also Turner, Pollmann and Berg) studied interaction
effects in one case: In their 1D time-reversal (TR) invariant topological
superconductor,\cite{FK1009,TPB1102,FK1103} the $\Z$ classification in the free
case breaks down to $\Z_8$ with interactions that preserve TR symmetry. Here we
present another model which illustrates the effects of interactions on 1D SPT
phases. We begin with a lattice Hamiltonian for a 1D superconductor with both
TR and $S_z$ spin-rotation symmetries, described by the $\Z$ classification in
the free case. With the addition of weak interactions that preserve these
symmetries, the classification reduces to $\Z_4$ (see Table \ref{tab:results}). 

\begin{table}
\centering
\begin{tabular}{ |c| c| c| }
\hline 
 Symmetry  & Free & With \\
 & classification & interactions\\\hline 
  $U(1)\times Z^T_2$ & $\Z$ & $\Z_4$ \\
 $Z_n \times Z^T_2$ ($n$ even) & $\Z$ & $\Z_4$ \\
$Z_n \times Z^T_2$  ($n$ odd and $n>1$)  & $\Z$ & $\Z_2$  \\\hline 
\end{tabular}
\caption{Symmetry groups described by the 1D Hamiltonian in Eq. 1 (where
$Z_2^T$ is generated by the time reversal transformation), with their free
fermion classification and how they reduce with interactions. The latter
remains true with strong interactions so all phases can be realized with free
fermions.}
\label{tab:results}
\end{table}

We compare these four fermionic phases to the four phases predicted separately
from group cohomology\cite{CGW1107,SPC1032,CGW1128} (a method valid for strong interactions). We find each
fermionic phase has a distinct projective representation\cite{PBT1039,CGW1107}
and since group cohomology also gives rise to four and only four distinct
phases,\cite{CGL1172} we conclude that all free fermions can
realize all strongly interacting SPT phases in this case. 
We further study interaction effects on a 1D superconductor with $Z_n$ discrete
$S_z$ spin rotation and TR symmetries. For this symmetry group, we find the SPT
phases are classified by $\Z_4$ when $n=$ even and $\Z_2$ when $n=$ odd. Again,
these results are separately obtained both from perturbing our fermionic
lattice Hamiltonian and from the group cohomology classification for strong
interactions -- showing that again, all the strongly interacting topological
phases can be realized by non-interacting fermions.

\emph{Free fermion lattice model} --- We write a 1D Hamiltonian with a trivial and two non-trivial phases
\begin{eqnarray}
 H&=&-t\sum_{\langle ij\rangle\sigma}c^\dag_{i\sigma}c_{j\sigma}-2\Delta_s\sum_jc^\dag_{j\up}c^\dag_{j\down} + \textrm{h.c.}\nonumber \\
&\pm&i\Delta_p/2\sum_jc^\dag_{j+1\up}c^\dag_{j\down}+c_{j+1\down}c_{j\up}+ \textrm{h.c.}
\end{eqnarray}
where the first term is typical nearest-neighbor hopping, the second term
$\Delta_s$ represents on-site pairing and the last term with $\Delta_p$ pairs
electrons on adjacent sites. 

\begin{figure}[tb]
\begin{center}
\includegraphics[width=0.9\linewidth]{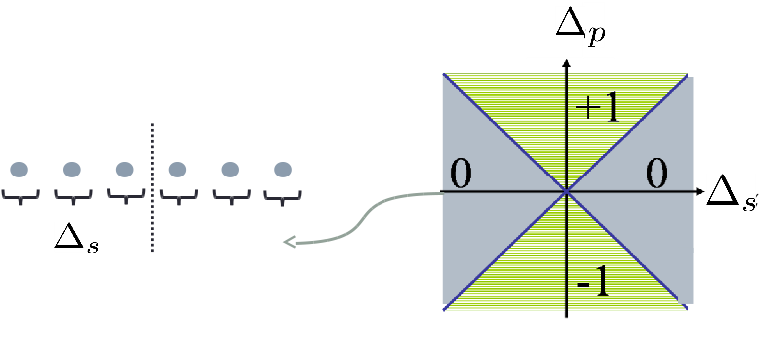}
\caption{Phase diagram when varying the parameters $\Delta_s$ and $\Delta_p$: The phase boundaries are $\Delta_s=\pm\Delta_p$ which separate three phases denoted by $N=+1, 0$ and -1. In the center phase we can make $\Delta_s$ arbitrarily large without closing the gap. That limit describes on-site pairing where any cut through the system separates both parts cleanly without leaving edge states -- allowing identification of this trivial $N=0$ phase.}\label{fig:phased}
\end{center}
\end{figure}

This Hamiltonian satisfies time-reversal $T$ and $S_z$ spin-rotation symmetries specified on $c_{i\sigma}^T=\{\hat{c}_{i\up},\hat{c}_{i\down}\}$ as
\[  \hat{T}c_{i\sigma}\hat{T}^{-1}=i\sigma_yc_{i\sigma};\quad e^{i\theta\hat{S}_z}c_{i\sigma}e^{-i\theta\hat{S}_z}=\left( \begin{array}{cc}
e^{-i\theta/2} &   \\
 & e^{i\theta/2} \end{array} \right)c_{i\sigma}\] 
so that
 $\hat{T}H\hat{T}^{-1}=H$ and $e^{i\theta\hat{S}_z}He^{-i\theta\hat{S}_z}=H$.
As the bandgap closes to leave just the hopping component
when $\Delta_s=\pm\Delta_p$, we obtain the phase diagram in Fig.
\ref{fig:phased}. 

We start by identifying the trivial phase $N=0$: When $|\Delta_s|>|\Delta_p|$, we can arbitrarily increase the strength of
$\Delta_s$ without closing the gap. In the
limit of $\Delta_s$ much larger than the other terms, the Hamiltonian simply reduces to on-site pairing --- any cut through the
system separates both parts cleanly leaving no boundary states
(see Fig. \ref{fig:phased}): this is the trivial $N=0$ phase. 

Next, we look for ground-state degeneracy at the interface between this phase and its
neighbors. This is most conveniently done in a low-energy continuum model, where the effective Hamiltonian becomes
\[ H=-i\int dx \tilde{\Psi}^\dag\left[(\sigma_z\otimes\mathbb{I} )\partial_x+\left( \begin{array}{cc}
& m   \\
-m^T & \end{array} \right)\right]\tilde{\Psi}\]
in a basis of right and left-moving fermion operators $\tilde{\Psi}^T=\big(\psi_{R\up},i\psi^\dag_{R\down},i\psi^\dag_{L\down},\psi_{L\up}\big)$ close to the Fermi surface. Here $m=\Delta_p\mathbb{I}-\Delta_s\sigma_z$.

Smoothly varying our mass term $m(x)$ across an interface, we set $\Delta_p(x)=\frac{1}{2}(1+\tanh x)$ and $\Delta_s(x)=\frac{1}{2}(1-\tanh x)$. 
This has the zero-energy solution
\begin{eqnarray}
 \hat{\psi}_{0+}&=&
\int dx\textrm{sech}(x)(\hat{\psi}_{R\up}+i\hat{\psi}^\dag_{L\down})\label{eq:localized}
\end{eqnarray}
This complex fermion operator ($\hat{\psi}_{0+}\neq \hat{\psi}^\dag_{0+}$) with energy $E=0$ contains a double degeneracy (empty or filled) that allows labelling of $\Delta_p>|\Delta_s|$ as the non-trivial $N=1$ phase. This mode transforms under symmetry as
\begin{eqnarray} \hat{T}\left( \begin{array}{c}
\hat{\psi}_{0+}  \\
\hat{\psi}_{0+}^\dag \end{array} \right)\hat{T}^{-1}&=&-\sigma_y\left( \begin{array}{c}
\hat{\psi}_{0+}  \\
\hat{\psi}_{0+}^\dag \end{array} \right),\nonumber\\
e^{i\theta\hat{S}_z}\hat{\psi}_{0+}e^{-i\theta\hat{S}_z}&=&e^{-i\theta/2}\hat{\psi}_{0+}\label{eq:zerosym}
\end{eqnarray}
Since the two degenerate states differ by $S_z=1/2$ and are related by time-reversal, each state carries quantum number of $S_z=\pm1/4$ respectively.

Using the symmetry relations in Eq. \ref{eq:zerosym}, we check if any perturbations in the Hamiltonian can shift the energy of this mode. We find density terms $\delta H = c\hat{\psi}_{0+}^\dag\hat{\psi}_{0+}$ are forbidden by TR, hence our ground-state degeneracy is protected by system symmetries --- this $N=1$ phase is stable against perturbations. 

To find the $N=-1$ phase, we change $\Delta_p(x)\to-\Delta_p(x)$ and upon repeating our procedure, find a different zero mode solution that we label
\begin{eqnarray}
\hat{\psi}_{0-}=\int dx\textrm{sech}(x)(i\hat{\psi}^\dag_{R\down}-\hat{\psi}_{L\up})
\end{eqnarray}
and instead transforms as 
\begin{eqnarray} 
\hat{T}\left( \begin{array}{c}
\hat{\psi}_{0-}  \\
\hat{\psi}_{0-}^\dag \end{array} \right)\hat{T}^{-1}&=&\sigma_y\left( \begin{array}{c}
\hat{\psi}_{0-}  \\
\hat{\psi}_{0-}^\dag \end{array} \right),\nonumber\\
e^{i\theta\hat{S}_z}\hat{\psi}_{0-}e^{-i\theta\hat{S}_z}&=&e^{-i\theta/2}\hat{\psi}_{0-}\label{eq:zerosymm}
\end{eqnarray}
This state has stable ground-state degeneracy as $\delta H=c\hat{\psi}_{0-}^\dag\hat{\psi}_{0-}$ is also forbidden by TR, indicating $\Delta_p<|\Delta_s|$ is a non-trivial phase as well.

Is it meaningful to label this second non-trivial phase $N=-1$? We examine what happens upon stacking two chains both with non-trivial phases but the first with $\Delta_p>|\Delta_s|$ and the second with $\Delta_p<|\Delta_s|$. Thus the first chain would have a zero mode $\hat{\psi}_{0+}$ and the second would have $\hat{\psi}_{0-}$. We find the coupling $\delta H = c\hat{\psi}_{0+}^\dag\hat{\psi}_{0-} +\textrm{h.c.}$ is allowed within system symmetries and makes the ground state nondegenerate. So two chains with the two distinct zero modes (labelled $+$ and $-$) combine to become trivial, indicating the two phases \emph{should} be labelled with opposite index. Naturally then the phase with $\hat{\psi}_{0-}$ would be the $N=-1$ phase, so 
this model indeed gives three symmetry protected phases $N=-1,0$ and +1.

While two chains containing zero modes with opposite index become trivial, we further consider the stability of two chains containing zero modes with the same positive (or negative) index. This may generalize to larger integers in the $\Z$ group, so we now examine the stacking of two chains with similar index more systematically. 

\begin{figure}[tb]
\begin{center}
\includegraphics[width=0.8\linewidth]{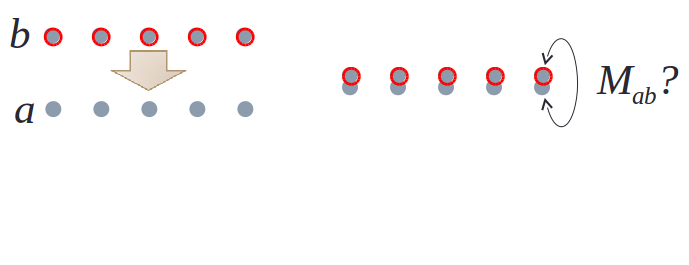}
\caption{We stack two chains in the same non-trivial phase with positive (or negative) index to see if their edge states are stable. With the first chain $a=1$ and the second $b=2$, $M_{ab}$ is any coupling between them. We find all possible couplings are forbidden by our system symmetries so two similar modes are stable and form the $N=2$ phase.}\label{fig:stack}
\end{center}
\end{figure}

A generic coupling term (see Fig. \ref{fig:stack}) is $\delta H=c\hat{\psi}_{+a}^\dag M_{ab}\hat{\psi}_{+b}+\textrm{h.c.}$. Here $a$ and $b$ are indices running over the chain number 1,2... e.g. $\hat{\psi}_{+1}$ denotes a zero mode from the $N=1$ phase in the first chain; and $M_{ab}$ is any generic coupling between these two operators. We examine the simplest case of $a=1$ and $b=2$. 
Terms of the form $\delta H=c\hat{\psi}_{+1}^\dag M_{12}\hat{\psi}_{+2}+\textrm{h.c.}$ are forbidden by TR symmetry as specified in Eq. \ref{eq:zerosym}, while fermion pairing terms such as $\delta H=c\hat{\psi}_{+1}^\dag M_{12}\hat{\psi}_{+2}^\dag+\textrm{h.c.}$ violate $S_z$ spin rotation symmetry. As there are no other quadratic fermion terms, the stacking of two chains is stable against perturbations and combine to give an $N=2$ phase.

Hence adding a number of 1D chains with positive index corresponds to a positive integer in the $\Z$ group. The negative numbers are obtained simply by stacking chains with $\hat{\psi}_{0-}$. As we showed earlier that a pair of $\hat{\psi}_{0+}$ and $\hat{\psi}_{0-}$ coupled together become trivial, the integer $N$ in our $\Z$ group is the difference between all positive and negative zero modes. Then for each phase labelled by $N$, the ground-state degeneracy is $2^{|N|}$.

\emph{Interaction effects} --- Now we allow couplings with an arbitrary
number of fermion operators. We look at terms with four and two operators which take the general form 
\begin{eqnarray} 
 \delta H &=&
V_{abcd}\hat{\psi}_{+a}^\dag\hat{\psi}_{+b} \hat{\psi}_{+c}^\dag\hat{\psi}_{+d} 
+W_{ab}\hat{\psi}_{+a}^\dag\hat{\psi}_{+b} 
+ \textrm{h.c.} 
\end{eqnarray} 
$\delta H$ is compatible with both TR and $S_z$ spin-rotation
symmetry when $V_{abcd}$ and $W_{ab}$ satisfy certain
conditions.

\begin{figure}[tb]
\begin{center}
\includegraphics[width=0.62\linewidth]{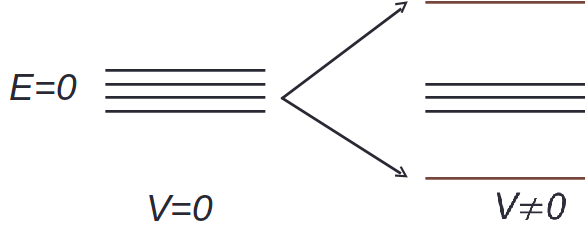}
\caption{Without interactions $V_{1234}=0$ the ground-state degeneracy for four zero modes is $2^4=16$. With interactions $V_{1234}\neq0$ two states are split by $\delta E=\pm|V_{1234}|$, making the ground state nondegenerate and this $N=4$ phase trivial. As $N=4$ is now smoothly connected to the trivial $N=0$ phase, our classification reduces from $\Z$ to $\Z_4$ with interactions.}\label{fig:esplit}
\end{center}
\end{figure}

A possible term couples four separate chains through an interaction with
only $V_{1234}\neq 0$.  Such an $\delta H$ is invariant under both TR
and $S_z$ spin-rotation symmetry.  This couples the two states $|0101\rangle$
and $|1010\rangle$ in our four-mode basis (0 and 1 denote unoccupied and
occupied respectively for each of the four chains). Without interactions we
have a ground-state degeneracy of $2^4=16$; with interactions two of these 16
states split in energy by $\delta E=\pm|V_{1234}|$, see Fig. \ref{fig:esplit}.
This makes the ground-state nondegenerate and the phase $N=4$ trivial. 

Since four chains with all positive (or negative) index are equivalent to the trivial phase, we can smoothly connect the $N=3$ phase to the $N=-1$ phase by adding four chains with all negative index. So with only three distinct non-trivial phases, the $\Z$ integer classification for free fermions reduces to $\Z_4$ in the presence of interactions.

Four-fermion interaction terms also reduce the ground-state degeneracy in the $N=2$ phase from $2^2=4$ to a two-fold degeneracy. The term \begin{eqnarray}
\delta H &=&V_{1122}(\hat{\psi}_{+1}^\dag\hat{\psi}_{+1}-\hat{\psi}_{+1}\hat{\psi}_{+1}^\dag)(\hat{\psi}_{+2}^\dag\hat{\psi}_{+2}-\hat{\psi}_{+2}\hat{\psi}_{+2}^\dag)\nonumber
\end{eqnarray}
causes two states $|00\rangle$ and $|11\rangle$ to shift in energy by $V_{1122}$ while two other states $|01\rangle$ and $|10\rangle$ shift by $-V_{1122}$. As we still have doubled ground-state degeneracy, the state $N=2$ remains non-trivial. To summarize, interaction effects reduce our degeneracy leaving three non-trivial phases each with a two-fold ground state degeneracy. 

\emph{Four distinct projective representations} --- Our results depend on perturbating from free fermions, i.e. we check the stability of free fermion phases with weak interactions. This method may not capture all possible interacting phases as strongly
interacting topological phases may not adiabatically connect to free
fermion phases.  Or, different phases from weak interactions may become the same phase with strong interactions. To address these issues, we calculate the projective representation for each phase to see if they are truly distinct. 

Using the symmetry operations defined in Eqs. \ref{eq:zerosym} and
\ref{eq:zerosymm}, we write their matrix representation on the degenerate
subspace. In the $N=1$ phase with basis $|0_+\rangle$ and $|1_+\rangle$
\[  U_\theta\to M(U_\theta)=\left( \begin{array}{cc}
1 &  \\
& e^{-i\theta/2} \end{array} \right),\quad \tilde{T}\to M(\tilde{T})K=\sigma_xK\]
where $\tilde{T}=U_{-\pi}T$, a rotated TR operator we can introduce since $U_\theta$ and $T$ commute. Then 
\begin{eqnarray}
M(\tilde{T})KM(U_\theta)=e^{i\theta/2}M(U_\theta)M(\tilde{T})K      \label{eq:pro1}
     \end{eqnarray}
and $M(\tilde{T})KM(\tilde{T})K=1$. This is a projective representation as the phase in Eq. \ref{eq:pro1} cannot be removed by adding any phase factor to $M(U_\theta)$ that changes the spin quantum number. 

Moving to the $N=-1$ phase with ground states $|0_-\rangle$ and $|1_-\rangle$, we have the same representation for $M(U_\theta)$ while $M(\tilde{T})K=e^{-i\sigma_z\pi/2}\sigma_xK$ in this case. Eq. \ref{eq:pro1} remains true but now $M(\tilde{T})KM(\tilde{T})K=-1$ so we have a different projective representation.

In the $N=2$ phase, our ground states are four-fold: $|0_+0_+\rangle, |0_+1_+\rangle, |1_+0_+\rangle$ and $|1_+1_+\rangle$. Here 
\begin{eqnarray}
M(U_\theta) &=& \left( \begin{array}{cc}
e^{i\theta/4} &\\
& e^{-i\theta/4} \end{array} \right)\otimes
\left( \begin{array}{cc}
e^{i\theta/4} &\\
& e^{-i\theta/4} \end{array} \right),
\nonumber\\
M(\tilde{T})K &=&\sigma_x\otimes\sigma_y
K.
\end{eqnarray}
While $M(\tilde{T})K$ and $M(U_\theta)$ commute this time, $M(\tilde{T})KM(\tilde{T})K=-1$ again making a third non-trivial projective representation.

As each non-trivial phase has a distinct non-trivial projective
representation, they remain distinct phases even when interactions are strong.
We can compare our results to the unperturbative bosonic classification in 1D
obtained by group cohomology\cite{CGW1107,CGW1128,CGL1172} as we can bosonize our fermionic model.
The resulting bosonic model would have the same symmetry $U(1)\times Z^T_2$ with phases
classified by $\Z_2\times\Z_2$, i.e. four distinct projective
representations there correspond to four different strongly interacting phases.
Our fermionic results  similarly contain all four phases with these distinct
projective representations, so this model realizes all possible
non-trivial phases with strong interactions. 

\renewcommand{\t}[1]{\tilde{#1}}
\newcommand{\e}{\hspace{1pt}\mathrm{e}}
\newcommand{\dd}{\hspace{1pt}\mathrm{d}}
\newcommand{\imth}{\hspace{1pt}\mathrm{i}\hspace{1pt}}

\emph{Modifying symmetry from $U(1)$ to $Z_n$ spin-rotation} --- As our fermion model respects $S_z$ spin-rotation and TR symmetry,
it naturally contains $Z_n$ discrete spin-rotation as well. We can replace
$U(1)$ spin-rotation by $Z_n$ spin-rotation, i.e. rotation by an arbitrary angle is now constrained to values of $\theta=2\pi/n$ and our new symmetry group has generators time-reversal $T$ and  discrete $S_z$ rotation $R=\e^{\imth S_z \frac{2\pi}{n}}$ satisfying 
\begin{align}
T^2=(-)^{N_F}, \ \ \ \ 
R^n=(-)^{N_F}, \ \ \ \
TR=RT.\label{eq:znsym}
\end{align}
Here $(-)^{N_F}$ is the fermion number parity operator. When $n=$ even, this group $G(T,Z_n)$ is generated by $R$ and $\t T=R^{n/2}T$, so $G(T,Z_n)=Z_{2n}\times Z_2^{\t T}$. When $n=$ odd, we find that $\t R=RT$ alone generates this group $G(T,Z_n)=Z_{4n}^T$\cite{supp}. 

For $n\geq2$, no new fermion bilinear terms are allowed so the free fermion classification does not change from $\Z$. In the case of $n=1$, new quadratic terms of the form  $\delta H=c\hat{\psi}_{+1}^\dag\hat{\psi}_{+2}^\dag + \textrm{h.c.}$ are permitted. This term couples two chains forming the $N=2$ phase to make the ground-state nondegenerate. The $N=2$ phase becomes trivial and the classification for $n=1$ reduces to $\Z_2$.

Similarly for higher $n$, we can always add interacting terms with $2n$ $\hat{\psi}_{+}$ operators similar to the term in the $n=1$ case above. For $n=2$ for instance, this term is $\delta H=c\hat{\psi}_{+1}^\dag\hat{\psi}_{+2}^\dag\hat{\psi}_{+3}^\dag\hat{\psi}_{+4}^\dag + \textrm{h.c.}$. Such interactions couple $2n$ zero modes each in the $N=1$ phase to render the ground-state nondegenerate. In effect, $Z_n$ spin-rotation symmetry allows interactions that reduce the classification to $\Z_{2n}$.

We had established that under $U(1)$ spin-rotation symmetry, interactions reduce the classification to $\Z_4$. Including more interactions as allowed by $Z_n$ spin-rotation further reduce the classification to $\Z_{2n}$. Taken together, we find there is no effect on even $n$ which remains $\Z_4$ since $2n$ is a multiple of 4. Odd $n$, however, reduces to a $\Z_2$ classification (as 2 becomes the largest common denominator between $2n$ and 4). 

The number of non-trivial phases can be compared to and matches with the group cohomology prediction $\Z_2\times\Z_2$ for even $n$ and $\Z_2$ for odd $n$\cite{supp}. We find that different symmetry groups with the same free fermion classification reduce to various results (here $\Z_4$ or $\Z_2$ are examples) in the presence of interactions (summary in Table \ref{tab:results}). As verified by comparison of these phases with group cohomology, all possible strongly-interacting phases can be realized by free fermions in this model.

Lastly, we note that our classification is protected only by system symmetries of spin-rotation and TR. As shown earlier, without such symmetry a term $\delta H = c\hat{\psi}_{0+}^\dag\hat{\psi}_{0+}$ would be permitted which renders the ground-state nondegenerate and the classification trivial ($\Z_1$).

\emph{Discussion} --- We study the SPT phases of 1D
fermionic superconductors with TR and $S_z$ spin-rotation
symmetries.  If fermions do not interact, their classification is given by the
$\Z$ group; with weak interactions this reduces to a $\Z_4$ classification. As each of our four fermion phases have distinct projective representations, they correspond to four distinct phases by comparison with group cohomology, which predicts four and only
four different gapped phases even with strong interactions. Hence all
distinct symmetric gapped phases with strong interactions are realized by non-interacting fermions in this case.

As the edge states in our 1D superconductors are described by
complex fermions which have double the degrees of freedom
of Majorana fermions, it is unsurprising that the interacting classification
is half of the result from Kitaev and Fidkowski's Majorana model\cite{FK1009,TPB1102,FK1103}. 

We further studied the  SPT phases of  1D superconductors with TR
and $Z_n$ discrete $S_z$ spin-rotation symmetries, to find they are
classified by $\Z_4$ when $n=$ even and $\Z_2$ when $n=$ odd.  Again, as phases
in our fermionic model matches with the group cohomology prediction, all gapped
phases of these 1D fermionic superconductors are also realized
by non-interacting fermions.

Interactions on different symmetry groups with the same free
fermion classification give rise to varied results (Table \ref{tab:results}). Here perturbing from a free fermion model gives all strongly interacting
phases, however in other cases such phases may not be realizable with free fermions. Lastly, the effectiveness of this method remains open especially in higher dimensions
where additional tools may be needed. Further study of different
symmetry groups or in higher dimensions would be worthwhile. 

Towards the completion of this paper, we noted the work of A.  Rosch
(arXiv:1203.5541) which shows ``a topological insulator made of four
chains of superconducting spinless fermions characterized by four Majorana edge
states can adiabatically be deformed into a trivial band insulator'' via
``interactions to {\em spinful} fermions'', which has some relation to our
$\Z_4$ classification of  1D fermionic superconducting phases with
TR and $S_z$ spin-rotation symmetries.

We thank Zheng-Xin Liu, Andrew Potter and Xie Chen for helpful
discussions. This work is supported by NSF Grant No. DMR-1005541 and NSFC
11074140.

\appendix
\section{Free fermion classification}
Given our many-body symmetries $T$ and $Q$ (where $Q$ denotes our U(1) symmetry), the mass matrix M should satisfy the following relations\cite{W1103}
\begin{eqnarray}
 &&\{\gamma,T\}=\{\gamma,M\}=\{T,M\}=0; \qquad M^2=-1\nonumber\\
&&[Q,\gamma]=[Q,T]=0\label{eq:conditions}
\end{eqnarray}
where $\gamma$ is simply the hopping part of our Hamiltonian.

It is simplest to choose a basis where $Q=I\otimes\varepsilon$ (here we write $\varepsilon=-i\sigma_y$) as the the anti-symmetric Hamiltonian we call $A$ takes the form\cite{W1103} 
\begin{eqnarray}
 A=H_a\otimes I+H_s\otimes\varepsilon; \qquad H=H_s+iH_a\label{eq:complex}
\end{eqnarray}
and $H_s=H^T_s$ and $H_a=-H^T_a$; i.e. the space simplifies into the complex space $C_0$. 

Given this choice our operators can take the form
\begin{eqnarray}
 \gamma&=&\sigma_z\otimes I\otimes I\nonumber\\
Q&=&I\otimes I\otimes\varepsilon\nonumber\\
T&=&\sigma_x\otimes I\otimes\varepsilon
\end{eqnarray}
which satisfy the conditions in Eq. \ref{eq:conditions}. Our Hamiltonian becomes the expression given in Eq. \ref{eq:complex} so we easily identify the space of our mass matrix:
\begin{eqnarray}
 M=\varepsilon\otimes \{I, \vect{n}\cdot\vect{\sigma},-I\}\otimes I
\end{eqnarray}
i.e. the three possible phases have eigenvalues of $+1, \pm1$ and $-1$ respectively. (Note that here $\vect{n}\cdot \sigma=n_x\sigma_x+n_y\varepsilon+n_z\sigma_z$ i.e. $M$ is real.) These three phases correspond to the $N=-1,0$ and $1$ phases discussed in the paper. 

Note the fermion number operator
$\hat{N}=\sum_i(\hat{c}^\dag_{i\uparrow}\hat{c}_{i\up} +
\hat{c}^\dag_{i\down}\hat{c}_{i\down})$ in this basis is
$N=\sigma_z\otimes\sigma_z\otimes\varepsilon$. As this does not commute with
the mass matrix in both the $\pm1$ phases, fermion number is not conserved ---
for our basis we choose $U(1)$ symmetry to represent spin conservation instead
of particle number conservation.

\section{Group cohomology classification}
Here we calculate the group cohomology for the symmetry group we denote as $G(T,Z_n)$, generated by the time reversal $T$ and 
 discrete $S_z$ rotation $R=\e^{\imth S_z \frac{2\pi}{n}}$.
The generators $T$ and $R$ satisfy
\begin{align}
T^2=(-)^{N_F}, \ \ \ \ 
R^n=(-)^{N_F}, \ \ \ \
TR=RT.
\end{align}
Here $(-)^{N_F}$ is the fermion number parity operator.

When $n=$ even, we see that $G(T,Z_n)$ is generated by
$R$ and $\t T=R^{n/2}T$, which satisfy
\begin{align}
 \t T^2=1, \ \ \ \ 
R^n=(-)^{N_F}, \ \ \ \
\t TR=R\t T.
\end{align}
Thus $G(T,Z_n)=Z_{2n}\times Z_2^{\t T}$ when $n=$ even.
This allows us to obtain\cite{CGL1172}
\begin{align}
 H^2[G(T,Z_n),U_T(1)]= \Z_2^2, \ \ \ \ n=\text{even}.
\end{align}

When $n=$ odd, let $\t R=RT$. We find that $\t R^n=
[(-)^{N_F}]^{1+\frac{n-1}{2}}T$ and $\t R^{2n}=(-)^{N_F}$. So a combination of
$\t R^n$ and $\t R^{2n}$ will give us $T$ and a combination of $T$ and $\t R$
will give us $R$.  Thus $\t R$ alone generates  $G(T,Z_n)$ when $n=$ odd.  We find that,
for odd $n$, $G(T,Z_n)=Z_{4n}^T$ which is generated by $\t R$ with $\t
R^{4n}=1$.  In $Z_{4n}^T$, $\t R^k$ is anti-unitary if $k=$ odd, and unitary if
$k=$ even.  We find $H^2[Z_{2n}^T,U_T(1)]= \Z_2$ (at least for
$n=1,2,3,4,5,6$).  Thus
\begin{align}
 H^2[G(T,Z_n),U_T(1)]= \Z_2, \ \ \ \ n=\text{odd}.
\end{align}
Thus the symmetry protected topological phases with time reversal and $Z_n$ discrete $S_z$ spin rotation
symmetries are classified by $\Z_2^2$ for even $n$ and $\Z_2$ for odd $n$.

%

\end{document}